\documentclass[10pt]{article}
\usepackage{authblk}
\usepackage[utf8]{inputenc}
\usepackage[english]{babel}
\usepackage{graphicx}
\usepackage[margin=0.8in]{geometry}
\usepackage[font=small]{caption}
\usepackage{xcolor}
\usepackage{amsmath,amssymb}
\usepackage[sort&compress, super, comma]{natbib}
\usepackage{float}
\usepackage[section]{placeins}
\title{Towards fully two-dimensional spintronic devices}

\author[1]{Alexey A. Kaverzin\thanks{corresponding author: a.kaverzin.rug@gmail.com}}
\author[1]{Talieh S. Ghiasi}
\author[2]{Avalon H. Dismukes}
\author[2]{Xavier Roy}
\author[1]{Bart J. van Wees}
\affil[1]{Zernike Institute for Advanced Materials, University of Groningen, Groningen, 9747 AG, The Netherlands}
\affil[2]{Department of Chemistry, Columbia University, New York, USA}

\begin{document}

\maketitle

\begin{abstract}
Within the field of spintronics major efforts are directed towards developing applications for spin-based transport devices made fully out of two-dimensional (2D) materials. In this work we present an experimental realization of a spin-valve device where the generation of the spin signal is exclusively attributed to the spin-dependent conductivity of the magnetic graphene resulting from the proximity of an interlayer antiferromagnet, CrSBr. We clearly demonstrate that the usage of the conventional 3D magnetic contacts, that are commonly air-sensitive and incompatible with practical technologies, can be fully avoided when graphene/CrSBr heterostructures are employed. Moreover, apart from providing exceptionally long spin relaxation length, the usage of graphene for both generation and transport of the spin allows to automatically avoid the conductivity mismatch between the source and the channel circuits that has to be considered when using conventional low-resistive contacts. Our results address a necessary step in the engineering of spintronic circuitry out of layered materials and precede further developments in the area of complex spin-logic devices. Moreover, we introduce a fabrication procedure where we designed and implemented a recipe for the preparation of electrodes \emph{via} a damage-free technique that offers an immediate advantage in the fields of air-sensitive and delicate organic materials.

\end{abstract}

\twocolumn
Giant magnetoresistance effect~\cite{baibich1988giant, binasch1989enhanced} and spin-transfer torque~\cite{slonczewski1996current, myers1999current} phenomena have already allowed for a breakthrough spin-based technology within the area of memory-related applications. Yet, the utilization of the spin degree of freedom within the scope of the semiconductor industry remains limited~\cite{awschalom2007challenges}. In order to progress further and make practical use of the spin transport functionality one has to advance substantially in every constituent of the spin transport devices. Fortunately, layered materials, being both rather versatile as a family~\cite{gong2017discovery, burch2018magnetism, gong2019two} and easy to assemble into a heterostructure~\cite{geim2013van}, offer a promising pathway to take in the view of both efficiency and size miniaturization and, thus, have become the main material choice for spintronic devices~\cite{cardoso2018van, zhai2021electrically, sierra2021van}. 

\begin{figure}[h]
    \centering
    \includegraphics[width=0.49\textwidth]{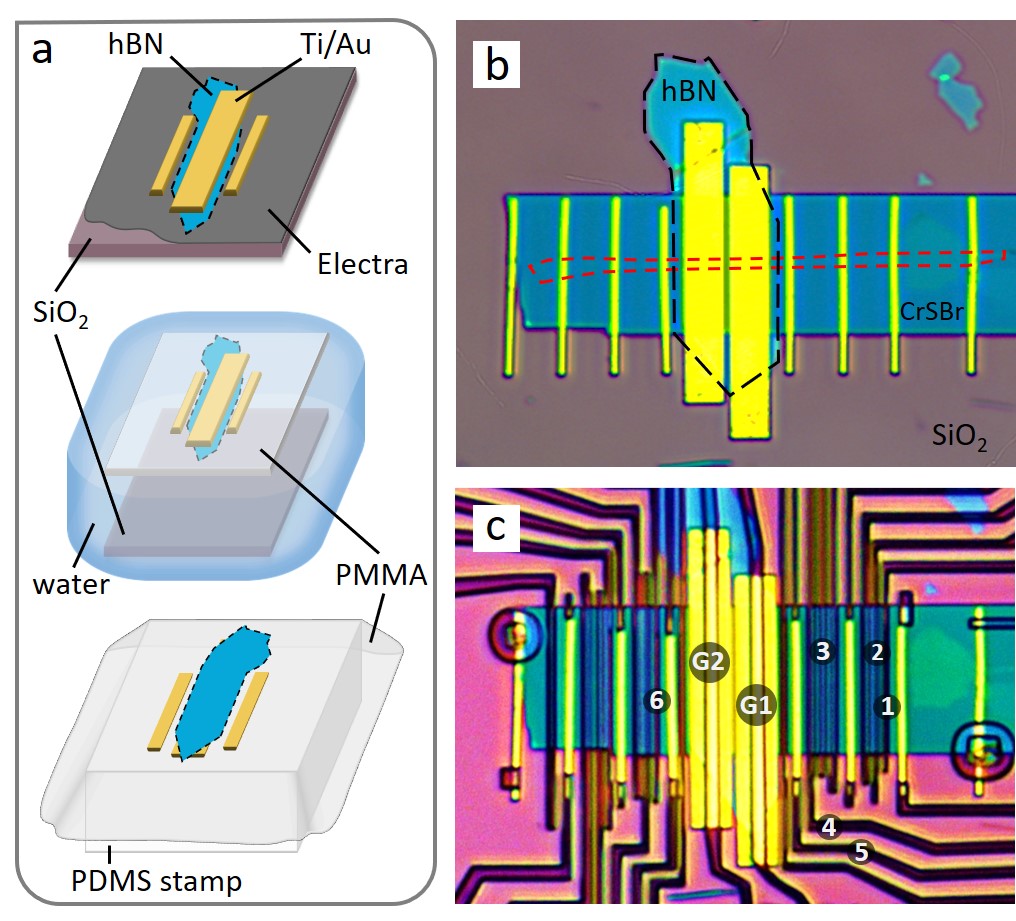}
    \caption{a. The developed process of the fabrication of stand-alone contacts is shown schematically. Top: Ti/Au contacts are fabricated \emph{via} regular lithography after exfoliation of hBN on Si/SiO$_2$/Electra substrate. After this step and before the step shown below the substrate is spin-coated with PMMA. Middle: PMMA layer with embedded in it Ti/Au contacts and hBN flake is suspended on water after Electra is dissolved. Bottom: PMMA is placed on top of a PDMS stamp and can be transferred later on a targeted substrate/flake with the help of a transfer stage. In our case it was used to both pick up bilayer graphene and transfer the resulting  structure onto a CrSBr flake. Panel b gives the sample image after the transfer is finalized. Graphene (hBN) is outlined with the red (black) dashed line. c. Optical image of the device after the development and before the final deposition of the TiOx/Co contacts.}
\end{figure}

\begin{figure*}[h]
    \centering
    \includegraphics[width=\textwidth]{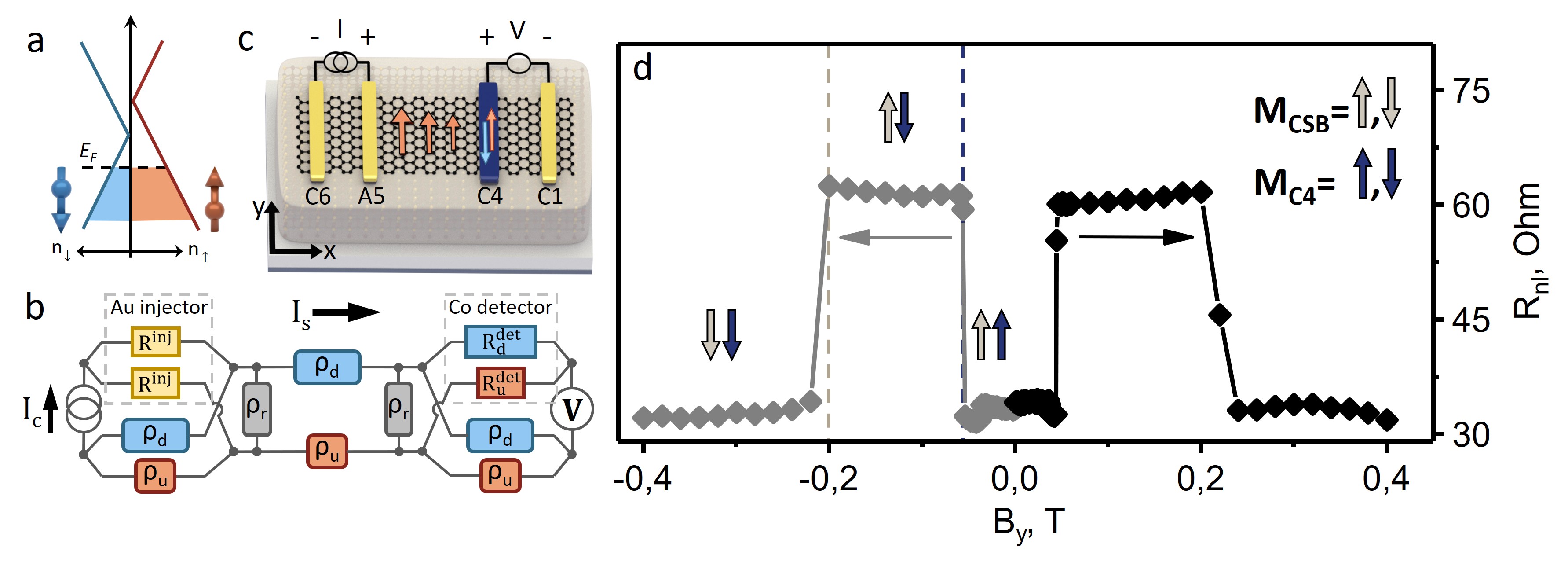}
    \caption{a. Exchange shift resulting in the spin polarization of the carrier density. b. Conventional schematics of the two channel model that is used to describe the transport in graphene with spin polarized conductance. Current is injected \emph{via} a non-magnetic contact, whereas the voltage is picked up \emph{via} a magnetic one. c. Schematics of the device together with circuit geometry. d. Spin valve measurement collected in the geometry schematically shown in the panel c having A5 as inner current injector and C4 as inner voltage detector. Black(grey) dots are measured when sweeping $B_x$ from 0$\,$T to 0.4$\,$T(-0.4$\,$T) with prior alignment done at -0.6$\,$T(0.6$\,$T). Grey and dark blue vertical arrows indicate the directions of $M_{CSB}$ and $M_{C4}$. Horizontal arrows indicate the sweeping direction of the applied magnetic field.}
\end{figure*}

Graphene is the prevailing host material for spin signals that can withstand the relaxation processes for a record long time and be transferred over tens of micrometer-long distance~\cite{Han2014,drogeler2016spin}. Yet graphene cannot offer the means for creation, manipulation and detection of the spins. Other 2D materials, however, can supplement what graphene lacks: when it is combined in a single heterostructure with an appropriately chosen companion it attains a spin-to-charge coupling \emph{via} the proximity effect that allows for an active generation and control of spin. Such possibility to combine the properties of different materials in a single structure has recently driven the booming interest in the van der Waals heterostructures~\cite{garcia2018spin, zollner2020swapping, sierra2021van, kurebayashi2022magnetism}. Of direct relevance for the spintronics are the reports that experimentally demonstrate spin Hall~\cite{safeer2019room}, Rashba-Edelstein~\cite{ghiasi2019charge, benitez2020tunable, khokhriakov2020gate}, Zeeman spin Hall~\cite{wei2016strong, behera2019proximity},  anomalous Hall~\cite{wang2015proximity, tang2018approaching, ghiasi2021electrical} and spin-dependent Seebeck~\cite{ghiasi2021electrical} effects, spin-dependent conductivity and other transport phenomena appeared/modified due to the present spin-orbit and/or exchange interactions induced in graphene~\cite{ghiasi2017large, benitez2018strongly, zihlmann2019large, ingla2021electrical, leutenantsmeyer2016proximity, singh2017strong, karpiak2019magnetic, wu2020large}.

Within this work we demonstrate the generation of the spin current exclusively by the graphene itself, which is possible when it is placed on top of a layered magnetic material such as chromium sulfide bromide (CrSBr). The CrSBr is an interlayer antiferromagnet with an inplane magnetic easy-axis~\cite{telford2020layered, lee2021magnetic}. As it was demonstrated in Ref~\cite{ghiasi2021electrical}, in a graphene/CrSBr heterostructure the large exchange shift of the band structure (estimated experimentally to be $\sim20\,$meV, Fig.$\,$2a) results in a considerable difference between the conductivities for the carriers of opposite spin alignment, \emph{i.e.} $\sigma_u\neq\sigma_d$. This directly implies a finite spin polarization of the graphene conductivity defined as $P_{Gr}=\frac{\sigma_u-\sigma_d}{\sigma_u+\sigma_d}$.

The finite $P_{Gr}$ grants graphene an active role in the generation and detection of the spin signal~\cite{zayets2012spin}. A charge current $I_c$, when passing through the graphene channel, generates an associated spin current $I_c P_{Gr}$ and vice versa. The initial experiments were so far performed on devices that had only spin-polarized electrodes which also inevitably contributed to the spin injection/detection~\cite{ghiasi2021electrical}. In contrast, here we prepare our stacks having both types of contacts (magnetic and non-magnetic) and demonstrate a distinct generation of the spin signal in graphene by using TiO$_\mathrm{x}$/Au electrodes only, without having the magnetic contacts involved in the spin current generation circuit. In this case the spin generation takes place exclusively within the graphene, which becomes a 2D source of the spin signal. The rough estimation of the spin polarization of graphene conductivity is found to be $\sim50\%$ and is of the same order of magnitude as the value reported in Ref.~\cite{ghiasi2021electrical}. 

For the preparation of the Au contacts we developed a novel recipe that allows less mechanical stress exerted on graphene during the pick-up procedure. In addition, it avoids multiple direct spin-coating of graphene with PMMA, which is known to introduce additional residues on the graphene surface. Our damage-free fabrication procedure has a great potential to be particularly beneficial for the materials that cannot withstand regular lithography-based preparation of metallic contacts while having similar flexibility and resolution. The newly developed part of the sample preparation is summarized in the schematics and sample images shown in Fig.$\,$1 and is explained in details in the Methods section. In short, we start with the exfoliation of hBN and fabrication of the Ti/Au contacts and gates (by e-beam lithography) on a water-soluble layer (Electra), as shown in panel a. By spin-coating a PMMA layer on top and dissolving the Electra in water, we release the hBN-Ti-Au-PMMA from the SiO$_2$ substrate. The PMMA layer floating on water is picked up and brought on top of a polydimethylsiloxane (PDMS) stamp and then can be transferred on any targeted flake on a separate substrate. In our case this hBN-Ti-Au-PMMA-PDMS structure was used to pick up a bilayer graphene flake and to transfer the resulting stack on top of a CrSBr flake. The resulting structure is shown in Fig.$\,$1b. Graphene and CrSBr flakes were exfoliated in advance on two separate Si/SiO$_2$ substrates. Panel c in Fig.$\,$1 gives the optical image of the sample taken just before depositing the magnetic electrodes made of AlO$_\mathrm{x}/$Co.

Overall, our observations confirm the robustness and consistency of the devices based on graphene/CrSBr stacks and offer an evident experimental demonstration of the generation of the spin current distinctly within a 2D system without the usage of conventional ferromagnetic contacts. Importantly, when graphene is used as both source and transport channel for the spin signal, the conductivity mismatch between the impedances of the source and transport circuits is automatically avoided which is not the case for the commonly used metallic contacts. Moreover, the spin relaxation length in magnetic graphene is much larger than that in the conventional ferromagnetic materials thus suggesting it as a more efficient source of the spin signal. Finally, being atomically thin, graphene allows an effective modulation of its Fermi level which in turn is expected to result in an active control of the spin valve action by the electric gate. All these facts promote graphene/CrSBr based devices as a very promising system for realising spin functionality in a fully 2D system where the spin action is controlled exclusively by electrical means.

\subsection*{Results and Discussion}

The graphene conductivity in the proximity of CrSBr becomes spin polarized due to the induced exchange splitting, Fig.$\,$2a. The presence of the exchange interaction implies a shift in energy between electron states polarized in ``up" and ``down" directions. We consider these ``up" and ``down" electrons as two separate carrier species of the current in our magnetic graphene which are dissociated from each other but can still communicate \emph{via} spin relaxation processes. Schematically such two-channel model can be depicted in an electrical circuit as shown in the panel b of the Fig.$\,$2, where $\rho_u$ and $\rho_d$ are corresponding resistivities in each of the two channels, $\rho_r$ (in units of $\Omega\cdot m$) is the resistivity that represents the connection between the two channels \emph{via} the present spin relaxation processes. Each electrical connection to the channel has to couple to both ``up" and ``down" channels \emph{via} the corresponding contact resistances. For the circuit shown in Fig.$\,$2b detection of the spin signal (right side of the circuit) is realized with the use of the magnetic material (Co). This implies that the contact resistances that couple the contact to the two spin channels are not equal, \emph{i.e.} $R^{det}_u\neq R^{det}_d$, and that the spin polarization of this contact resistance $P_d$ is nonzero. Contrary to the common implementation of the spin generation circuit, for the injection of the current here we use a non-magnetic electrode. In this case the spin polarization of the contact resistances $P_i$ is exactly zero since $R^{inj}_u=R^{inj}_d=R^{inj}$. Nonetheless, since the spin polarization of the channel resistance is finite, the spin current is still generated by graphene itself.

The mechanism of generation of the spin accumulation in magnetic graphene can be understood as follows: in a homogeneous magnetic system the charge current is spin polarised, since a larger portion of the current flows through the channel with the lowest resistivity ($\rho_u$ for the case shown in Fig.$\,$2a). When the charge current is passed through the left part of the channel only, the created discontinuity of the spin current at the injection point generates a spin accumulation. From there, the spin accumulation decays exponentially with the characteristic length scale $\lambda$. Considering the circuit shown in Fig.$\,$2b, it is also apparent that the injection circuit is a Wheatstone bridge. Passing the current through it will result in a voltage difference (spin accumulation) appearing between the ``up" and ``down" channels when the bridge is unbalanced.

\begin{figure*}[h]
    \centering
    \includegraphics[width=\textwidth]{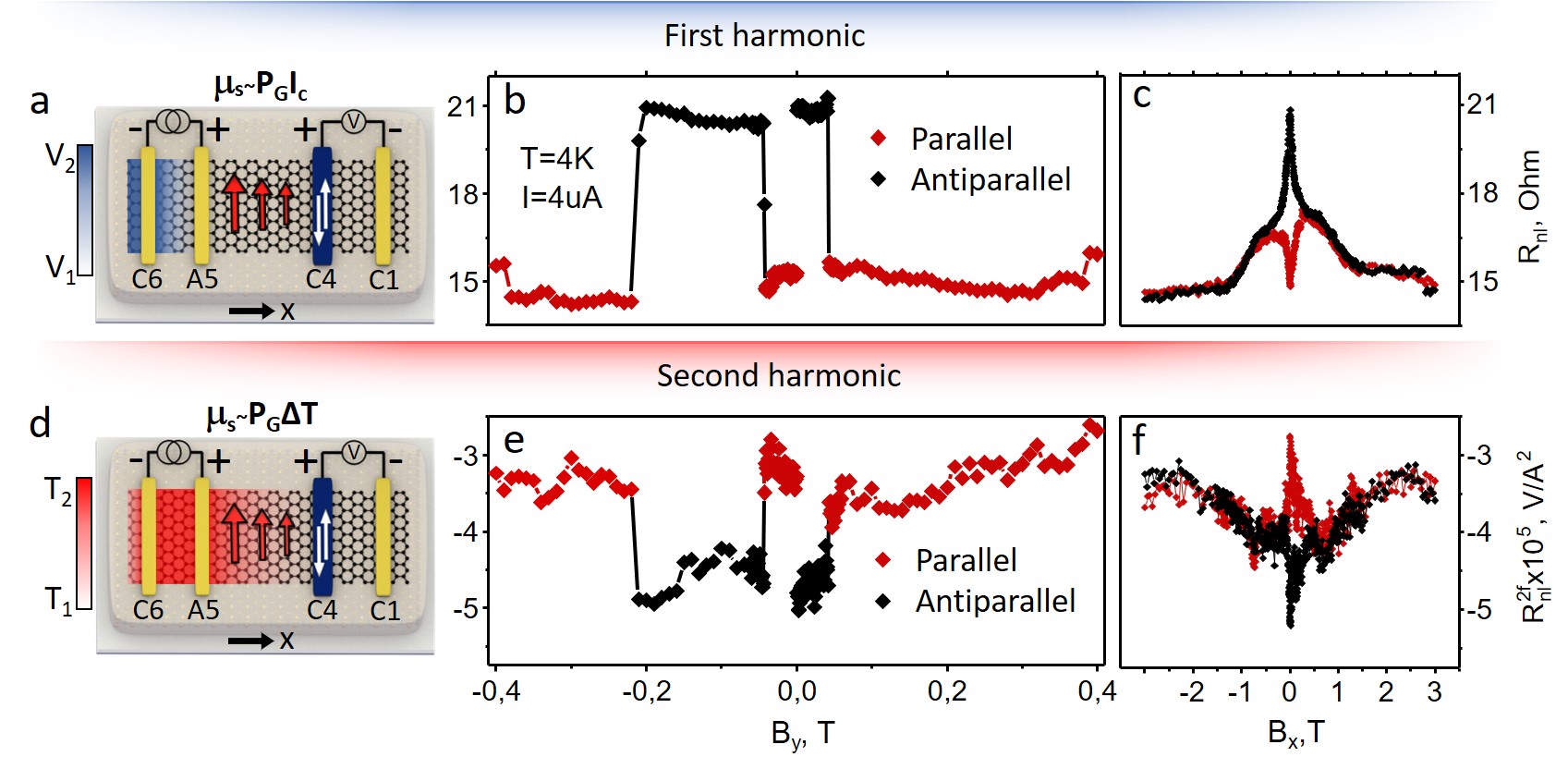}
    \caption{ a. Schematics of the sample circuit for the 1st harmonic measurements shown in panels b and c. Color scale represents the voltage distribution in the channel. d. Schematics of the sample circuit for the 2nd harmonic measurements shown in panels e and f. Color scale represents the temperature distribution in the channel. Measurements in spin valve (b and e) and Hanle (c and f) geometries collected at 1st (a and c) and 2nd (b and d) harmonics of the lock-in amplifier. Black and red dots represent parallel and anti-parallel alignment of $M_{CSB}$ and $M_{C4}$.}
\end{figure*}

The described above spin generation mechanism is very similar to a conventional one when the charge current is passed between a ferromagnetic material and a non-magnetic one~\cite{johnson1985interfacial}, yet it does not require an additional 3D ferromagnetic electrode. Similarity between these mechanisms is clear from our resistance model (Fig.$\,$2b) and is also directly reflected in the appearance of the relevant terms in the expression for the associated non-local resistance $R_{nl}$ derived in Ref.~\cite{ghiasi2021electrical}:
\begin{equation}
   R_{nl}=\frac{\lambda R_{sq}}{2W(1 - P^2_{Gr})}e^{-L/\lambda}(P_i - P_{Gr})(P_d - P_{Gr})
   \label{Eq:Rnl01}
\end{equation}
Here $W$, $R_{sq}$ and $\lambda$ are width, square resistance and spin relaxation length of the channel, respectively. The formula is derived for the case when both injecting and detecting contacts have a finite $P_i$ and $P_d$. The first pair of the parenthesis represents the total injection efficiency of the circuit where $P_{gr}$ and ${P_i}$ enter in a very similar fashion.

\begin{figure*}[h]
    \centering
    \includegraphics[width=\textwidth]{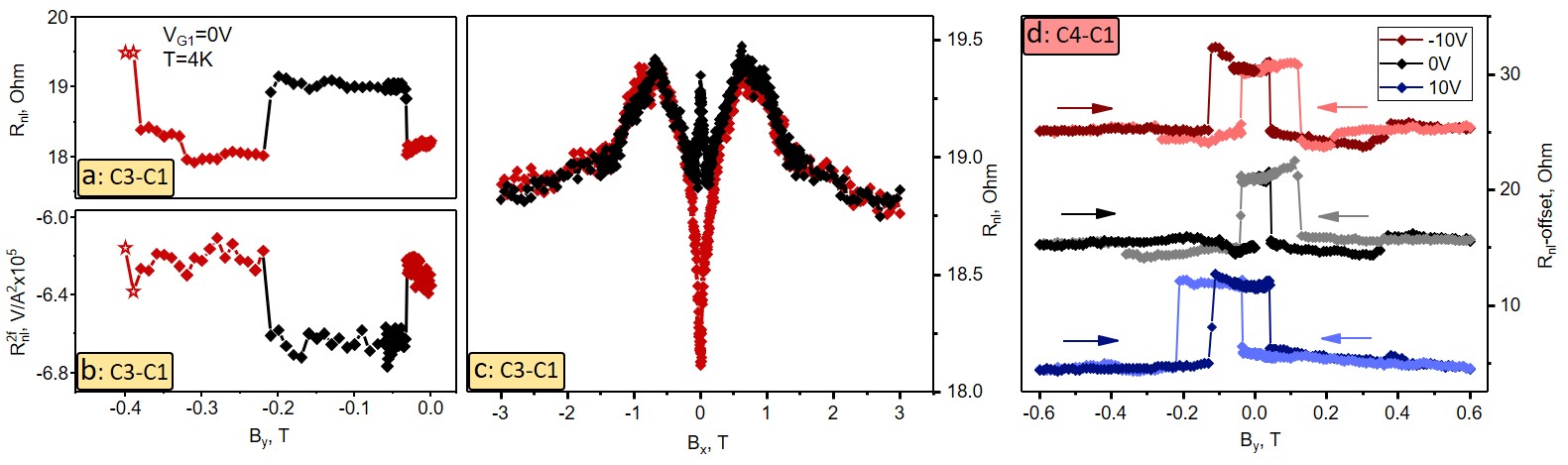}
    \caption{First (a) and second (b) harmonics spin valve measured at the second pair contacts C3-C1. Only left half is shown. Star-shaped data points are interpreted as being measured in the same magnetization alignment as other red color points, but in a different resistance state of the parallel conductance channel. c. Non-local measurement in Hanle geometry corresponding to panel a. Black and red dots represent parallel and anti-parallel alignment of $M_{CSB}$ and $M_{C3}$. d Spin valve measured at pair C4-C1 plotted for different applied gate voltages. Additional switch at $\pm0.38\,$T associated with the parallel conductive channel is suppressed at $V_{G1}=10\,$V. Arrows indicate the direction of the magnetic field sweeping.}
\end{figure*}

The second parenthesis in Eq.$\,$\ref{Eq:Rnl01} represents the reciprocal process in the spin circuitry, \emph{i.e.} spin detection. There are two components related to the spin polarization of the magnetic graphene and to that of the magnetic detector electrode. The two associated detection mechanisms enter the equation in a similar way, which is also reflected by the resistance circuit. 

In the experiment described here we inject the current \emph{via} a non-magnetic electrode (Fig.$\,$2c). Thus, $P_i=0$, which modifies the formula for the measured non-local signal as $R_{nl}\propto (P_i - P_{Gr})(P_d - P_{Gr}) {|_{P_i=0}=P_{Gr}(P_{Gr} - P_d)}$ resulting in two contributions. The first one is proportional to $P^2_{Gr}$ and gives a positive spin-related background which does not depend on the relative orientation of the magnetization of the top most layer of CrSBr, $M_{CSB}$, with respect to the magnetization of the detector. The second term is proportional to the product $P_{Gr}\cdot P_d$ and results in the two levels of the non-local resistance depending on the relative alignment of the magnetizations. This is exactly confirmed with the experiment shown in Fig.$\,$2d where the non-local spin valve is measured as a voltage difference between cobalt contacts C4 and C1 when the AC current is supplied between the main gold injector A5 and a reference cobalt contact C6 (contacts are numbered in Fig.$\,$1c). The measured voltage difference is normalized by the applied current which gives the non-local resistance $R_{nl}$. The measurement is performed as a function of the external magnetic field $B_y$ applied along the easy magnetic axis of both cobalt electrodes and CrSBr flake ($y$-axis, also referred to as crystallographic $b$-axis of CrSBr \cite{telford2020layered, lee2021magnetic}). The value of the measured non-local resistance is observed to have two clear levels as it is also expected from the derivation. In the case of a regular graphene channel on a SiO$_2$ substrate the same measurement would show no modulation of the non-local resistance since the injection current \emph{via} a gold contact has zero spin polarization. Yet, once graphene is in the proximity to a ferromagnetic substrate, the inherited exchange interaction results in graphene acting as a source of the spin accumulation detected by the magnetic contact. 

Note that in a homogeneously proximitized graphene channel, when only non-magnetic contacts are used, the separation of the spin-associated signal from the commonly present background (due to a non-ideal Ohmic current) is not straightforward. In that case the non-local spin signal would be independent of the direction of the graphene magnetization since both injection and detection circuits would share exactly the same magnetization determined by the underlying CrSBr. Therefore, we use spin-sensitive magnetic electrodes in order to unambiguously identify the spin currents generated when non-magnetic TiOx/Au contacts are employed in the injection circuit.

From Fig.$\,$2d we observe that the parallel alignment of $M_{CSB}$ and $M_{C4}$ gives a lower non-local resistance than the anti-parallel alignment. Therefore, based on the derived relation $R_{nl}\propto P_{Gr}(P_{Gr} - P_d)$ we conclude that $P_{Gr}$ is of the opposite sign compared to $P_{C4}$. Under the same assumptions as described in Ref.~\cite{ghiasi2021electrical} this may suggest that the graphene sample is hole doped which is in agreement with the dependence of graphene resistance on the applied $V_{G1}$ (see SI, sec. 2). However, since we are not able to tune the Fermi level across the Dirac point in the graphene channel between contact A5 and C4 we cannot reliably determine the position of the Fermi level and, thus, the size of the induced exchange splitting. Nevertheless, assuming that spin polarizations of both graphene conductivity and cobalt contacts resistance are equal, we are able to estimate the spin relaxation length to be $\sim450\,$nm (SI sec.$\,$4). 
This results in a rough estimate of the involved spin polarizations $P_{Gr}=-P_{C4}\approx50\,\%$. The possible uncertainty in the estimation of $P_{Gr}$ can be due to several factors, \emph{e.g.} inhomogeneity of the channel doping, non-equal spin polarization of the contact and graphene resistances, \emph{etc.} In particular, the uncertainty in the spin-relaxation length estimate is the most probable reason for the rather large calculated spin polarization of graphene conductivity. Nevertheless, the obtained number still suggests a consistently high efficiency of the spin signal generation within the magnetic graphene, thus, implying that in our spin transport circuit we can substitute the commonly used magnetic contacts with the regular non-magnetic ones and still obtain a signal of a similar magnitude.

In order to explore the possibility to tune/change the sign of the spin polarization of graphene conductivity we added two top-gates G1 and G2 into the design of the device (Fig.$\,$1c). In the full possible range of $[-10;10]\,$V applied to G1 we observe no significant change in the size of the switches yet there is a modulation of the background level (SI sec.$\,$5, Fig.$\,$4d) which is likely to be related to the change in the charge related background. Unfortunately, during the sweep of the gate G2, one of the side connections to the electrode A5 was lost and there was a significant change in the channel resistance covered by the hBN (SI sec.$\,$2). After the sample change the same spin valve geometry as used for Fig$\,$2d still shows a characteristic spin valve measurement, yet the magnitude is decreased from $\sim30\,\Omega$ to $\sim6.6\,\Omega$ as seen in Fig.$\,$3b. From now on we will be discussing the sample in the altered state. Note that the estimation of the spin relaxation length is done for the changed state of the sample since the distance dependence of the signal was measured only after the sample changed. Therefore, the calculation of the spin polarisarion of graphene conductivity described earlier is done using the spin valve measurement shown in Fig.$\,$3b where the size of the switch is $6.6\,\Omega$.

Furthermore, we noticed that the exact switching behavior of M$_{CSB}$ direction is not always reproducible and does depend on a value of the magnetic field used for the alignment. This in fact may be expected considering that in our measurements we are mostly sensitive to the magnetization of the top most layer of the bulk CrSBr whereas the full magnetization behavior is determined by the anti-ferromagnetic interaction between the top most and its neighboring layers. In Fig.$\,$3b we plot a spin valve measured with the same connections where for the positive range of the applied magnetic field the value of the $R_{nl}$ corresponds to an anti-parallel alignment. This implies that $M_{CSB}$ switched back towards positive direction of $y-$axis while the magnetic field was brought to zero after alignment at $-0.6\,$T. As a consequence with further increase of the field only the switch of $M_{C4}$ is observed at $44\,$mT after which $M_{CSB}$ and $M_{C4}$ remain in a parallel alignment.

To complete the investigation of the induced spin accumulation we supplement spin valve measurements with the R$_{nl}$ dependence on the magnetic field applied perpendicular to the easy axis (along $x$-axis), Fig.$\,$3c. Black and red curves correspond to parallel and anti-parallel alignments between $M_{CSB}$ and $M_{C4}$, respectively. Applying the magnetic field perpendicular to the alignment of the injected spins usually results in a Hanle precession of the spins that takes place while they diffuse from the injector to the detector contact. Yet, in magnetic graphene the induced exchange field $B_{exch}$ is strong enough to destroy all the spin components except those that are (anti)parallel to the direction of $B_{exch}$ irrespective of relatively weak applied external field. Under this condition spins are always aligned with the exchange field. The role of the external field is to change the directions of both the exchange field (parallel to $M_{CSB}$) and of the detecting electrode magnetization. Here and below we define the geometry where the non-local signal is measured with the magnetic field applied along $x-$axis as ``Hanle" geometry yet this is not a Hanle precession measurement as understood conventionally. 

The functional dependence of the change observed in $R_{nl}$ with applied $B_x$ is proportional to the cosine of the angle between $M_{CSB}$ and $M_{C4}$. Cobalt contacts are much softer magnetically compared to CrSBr along the $x$-axis. Based on SQUID measurements~\cite{telford2020layered}, within the magnetic field range of $B_x<0.2\,$T the direction of $M_{CSB}$ changes only by a few degrees while $M_{C4}$ becomes fully aligned with the field at $B_x\simeq0.2\,$T. Therefore, irrespective of the initial alignment between $M_{C4}$ and $M_{CSB}$ at $B_x\simeq0.2\,$T both black and red dotted curves merge and continue jointly at higher fields until the direction of $M_{CSB}$ saturates along the y-axis at $B_x\simeq1.4\,$T. Above this value both cobalt contact magnetization and injected spin direction coincide again and therefore the signal recovers its initial value at $B_x=0\,$T for the parallel alignment.

Together with the first harmonic response of the lock-in amplifier we collected the second harmonic that is commonly associated with the phenomena driven by the temperature gradient in the graphene channel induced by Joule heating. In Figs.$\,$3e,f both the second harmonic spin valve and dependence of the non-local signal on $B_x$ are shown, measured at the same time as those given in panels b and c. Similar to the first harmonic response there are 2 distinct levels of $R^{2f}_{nl}$ depending on the relative alignment of $M_{C4}$ and $M_{CSB}$, as expected. These observations clearly identify the spin origin of the measured signal and confirm its attribution to the spin-dependent Seebeck effect which results from the induced exchange interaction as in Ref.~\cite{ghiasi2021electrical}. Charge current generates a Joule heating which, due to the present spin-dependent Seebeck effect, results in a finite spin current and spin accumulation that is sensed by the ferromagnetic detector circuit. In Ref.~\cite{ghiasi2021electrical} it was concluded that the sign of the second harmonic signal does not depend on the position of the Fermi level which means that the switching behavior has to be identical throughout different samples irrespective of the doping level. This is experimentally justified here since the sign of the switch is the same as that reported in Ref.~\cite{ghiasi2021electrical}, thus, further confirming the consistency of our interpretation of the results. 

In Fig.$\,$3b next to the main $\sim6.6\,\Omega$ switch we observe a much smaller one of $\sim1\,\Omega$ in size which occurs at $B_y=\pm0.38\,$T (not visible in the second harmonic signal). In order to understand the origin of this additional switch, we compare the signal collected at the contacts pair C4-C1 with that collected at the pair C3-C1. Similar to C4, contact C3 is made out of cobalt but is placed further away at a distance of $1.9\,\mu$m from the Au injector A5 (center-to-center). The corresponding spin valve (1$^{st}$ and 2$^{nd}$ harmonic) and the dependence on $B_x$ (1$^{st}$ harmonic) are displayed in Figs.$\,$4a-c. First of all, the two switches occurring at $B_y=-32\,$mT and $B_y=-0.21\,$T are clearly associated with spin-related signal as it fully complies with the expected behavior. Namely, the non-local resistance switches up at the moment of the switch of the magnetization of contact C3 and it switches down together with the switch of $M_{CSB}$. Furthermore, when comparing the pairs C4-C1 and C3-C1 we find that the size of the spin signal decreases substantially with the distance between the injector and detector electrodes. This is fully in accordance with the expected change of the spin signal due to the spin relaxation processes.

Significantly, the magnitude of the additional switch is almost the same as for pair C4-C1 clearly indicating minimal scaling with the distance. As seen, its size is comparable with the spin-associated switch which implies that not only the spin valve but also the dependence on $B_x$ should be largely affected by this additional contribution. Indeed, as seen in Fig.$\,$4c the shape of the dependence is quite different from what is expected and observed for example for the pair C4-C1 (Fig.$\,$3c). Specifically one could point out that at high enough field $B_x\gtrsim2\,$T, when all the magnetizations are assumed to be aligned with the field, the value of the non-local resistance does not saturate at the same level as $R_{nl}(B_x=0)$. The distinction between the spin-related and the additional contributions is further accentuated by the apparent absence of the former one in the second harmonic spin valve, Fig.$\,$3b.

All these observations and particularly the absence of the signal scaling with the distance hint at the origin of the additional contribution to be associated with the charge transport through a parallel channel. In fact, CrSBr is a semiconductor that can have a finite conductivity due to the residual doping, although it is expected to be much lower than conductivity of graphene \cite{telford2020layered} (also see SI). The dependence of CrSBr resistance on the applied magnetic field (also in \cite{telford2020layered} and SI) further indicates that the behavior shown in Figs.$\,$4a,c is likely to be a combination of the charge related contribution associated with the CrSBr magnetoresistance and the spin-related contribution originating from the spin transport in graphene.
Charge transport in CrSBr is expected to be controlled by the position of the Fermi level that is tuned by applying a gate voltage. Therefore, we studied how the spin valve measurements on contacts pair C4-C1 changes when the voltage is applied on the local gate $G1$ within the range $[-10;10]\,$V. In Fig.$\,$4d the corresponding spin valves are plotted with an offset for clarity. Evidently the spin-associated signal does not change significantly under applying the gate voltage whereas the additional contribution does get suppressed at $V_{G1}=10\,$V. This can be understood assuming that by applying positive gate voltage we shift the Fermi level in CrSBr more into the band gap thus reducing its conductivity and blocking the unwanted parallel conduction channel. Thus, we conclude that the observed additional contribution to the measured signal is likely to be due to the finite resistivity of the CrSBr, yet we show that it is possible to distinguish it from the spin-related component by studying the dependence of it on the distance and/or gate voltage. 

\subsection*{Conclusions}

We have performed the non-local measurements in graphene/CrSBr heterostructure in both first and second harmonic and in both spin valve and Hanle geometries. We have demonstrated that by using exclusively a non-magnetic electrode we are able to create a finite spin accumulation inside the magnetic graphene. Moreover, the usage of graphene/CrSBr heterostructure for spin injection/transport offers other immediate advantages compared to conventional 3D magnetic electrodes. Firstly, the conductivity mismatch between the spin source and transport channel is automatically avoided which simplifies the optimisation of the performance of the spintronic circuit. Secondly, a large spin relaxation length in graphene suggests higher efficiency of magnetic graphene as a source of the spin signal in comparison to the conventional ferromagnetic contacts. Finally, a large Seebeck coefficient of graphene ensures the presence of the spin-dependent Seebeck effect and offers even richer functionality with coupling the spin and heat currents. Overall, our findings confirm the graphene/CrSBr heterostructure as a robust platform for studying spin transport in a 2D magnetic channel. The distinct generation of spin signal within a 2D system when using nonmagnetic electrodes together with the potential tunability of the spin valve action by an electrical gate introduces graphene/CrSBr-based devices as a technologically relevant block for building fully 2D spintronic/spin-caloritronic devices. In addition, we have developed and implemented in the measured devices a novel damage-free recipe for the preparation of the contacts separately from the studied flake/material. Such recipe is of a great value for air sensitive as well as for organic materials.

\subsection*{Methods and sample fabrication}
Devices D1-D3 were prepared starting with exfoliation of the necessary layered components, namely graphene, CrSBr and hBN. Graphene and CrSBr are exfoliated on top of doped Si/SiO$_2$ substrates. The details of the growth of CrSBr bulk crystals are given in Refs.$\,$\cite{ghiasi2021electrical,telford2020layered}. The flakes with appropriate thicknesses are identified by their optical contrast with respect to the SiO$_2$/Si substrate. For D1 (discussed in the manuscript) we used bilayer graphene and 20-40$\,$nm thick CrSBr flakes. The hBN flake is exfoliated on a separate Si/SiO$_2$ substrate that is preliminary covered with a water soluble conductive polymer Electra [AR-PC 5090.02, Allresist]. Electra is spin-coated with the rate of 1000$\,$rpm and baked afterwards at 95$\,^{\circ}$C for 1$\,$min on the hot plate. The thickness of the resulting Electra layer is about 200$\,$nm. When an appropriate hBN flake of 20-50 nm thickness is selected (that is intended to be an insulator for the top gate), a 500$\,$nm PMMA ($4\,\%$, 950$\,$K) layer is spin-coated on top of the Si/SiO$_2$/Electra/hBN substrate at a rate of 1000$\,$rpm. A freshly covered substrate is baked on the hot plate at 180$\,^{\circ}$C for 1$\,$min. By means of e-beam lithography an appropriately designed structure is exposed in PMMA, including contacts and the top gate electrodes. After developing, the substrate is loaded into an e-beam evaporation setup where $0.5\,$nm of Ti and $90\,$nm of Au is deposited. Lift-off is done in acetone either at room or elevated temperature of 45$\,^{\circ}$C and results in the structure schematically shown in Fig.$\,$1a (top). We use here the fact that Electra does not get dissolved in either acetone nor developer solution and stays intact.

Resulting Si/SiO$_2$/Electra/hBN/Ti/Au is again covered with PMMA using the same coating parameters as described earlier. In the next step the sample is attached to a scotch tape that has a 7 by 7 mm window centered at the hBN flake and then is immersed in water. Water dissolves Electra which leads to a gradual detachment of the Si/SiO$_2$ substrate that eventually sinks down while hydrophobic PMMA film stays floating on top of water together with the attached scotch tape, Fig.$\,$1a (middle, the scotch tape is not shown in the schematics). The tape with the PMMA film is taken from water, dried in air and later put on top of a PDMS stamp with an Au/Ti/hBN structure facing outwards, Fig.$\,$1a (bottom). At this stage the Ti layer is directly exposed to air and water and, thus, gets oxidized. Subsequently, the stamp with the PMMA film is used to pick up a chosen bilayer graphene flake from a separate Si/SiO$_2$ substrate. The pick-up surface of the full stamp is flat since both hBN layer and TiOx/Au contacts are fully imbedded in the PMMA film. Graphene/hBN/Ti(TiOx)/Au/PMMA heterostructure is thereafter placed onto a targeted CrSBr flake. The final stack is shown in Fig.$\,$1b where it is still covered with the PMMA layer. The same PMMA is used later on for the lithography and deposition of Al$_2$O$_3$/Co contacts. Picture of device D1 after development and just before the deposition of Al$_2$O$_3$/Co contacts is given in Fig.$\,$1c.

Our recipe offers several advantages from the perspective of fabrication of graphene-based samples that require different types of contacts simultaneously. First of all, before the pick-up of graphene flake, the hBN flake together with the Ti/Au contacts on Si/SiO$_2$/Electra substrate is covered with a fully relaxed film of PMMA. This assures a smooth gap-less pick-up surface of the prepared mask. Second of all, the Ti/Au contacts are made in advance on a separate substrate, thus, fully avoiding the risks of any errors occurring during this procedure. Third of all, the graphene is covered by the polymer only once (during the pick-up). Normally, during the preparation of the graphene-based structure with the two different types of contacts the sample (including graphene) is covered by the polymer 3 times: once during the stacking (when it is done using the polycarbonate film) and two times for fabricating separately the two different contact types. Reducing the number of steps where graphene is covered by the polymer is expected to lower the amount of residues on its surface.

Overall we have measured 3 devices where we were able to see the switching behavior of the non-local resistance associated with the spin signal where non-magnetic contacts were used either as the current injector or as the voltage detector. Further details of the fabrication and measurements on the other samples are given in the supplementary information.

\subsection*{Acknowledgements}
We would like to thank T. J. Schouten, H. Adema, A. Joshua, H. de Vries and J. G. Holstein for technical support. The presented research was funded by the Dutch Foundation for Fundamental Research on Matter (FOM) as a part of the Netherlands Organisation for Scientific Research (NWO), the European Union's Horizon 2020 research and innovation program under grant agreements No 696656 and 785219 (Graphene Flagship Core 2 and Core 3), NanoNed, the Zernike Institute for Advanced Materials, and the Spinoza Prize awarded in 2016 to B. J. van Wees by NWO. Synthesis, structural characterization and magnetic measurements received support as part of Programmable Quantum Materials, an Energy Frontier Research Center funded by the U.S. Department of Energy (DOE), Office of Science, Basic Energy Sciences (BES), under award DE-SC0019443. AD is supported by the NSF graduate research fellowship program (DGE 16-44869).

\providecommand{\latin}[1]{#1}
\makeatletter
\providecommand{\doi}
  {\begingroup\let\do\@makeother\dospecials
  \catcode`\{=1 \catcode`\}=2 \doi@aux}
\providecommand{\doi@aux}[1]{\endgroup\texttt{#1}}
\makeatother
\providecommand*\mcitethebibliography{\thebibliography}
\csname @ifundefined\endcsname{endmcitethebibliography}
  {\let\endmcitethebibliography\endthebibliography}{}


\begin{mcitethebibliography}{38}
\providecommand*\natexlab[1]{#1}
\providecommand*\mciteSetBstSublistMode[1]{}
\providecommand*\mciteSetBstMaxWidthForm[2]{}
\providecommand*\mciteBstWouldAddEndPuncttrue
  {\def\EndOfBibitem{\unskip.}}
\providecommand*\mciteBstWouldAddEndPunctfalse
  {\let\EndOfBibitem\relax}
\providecommand*\mciteSetBstMidEndSepPunct[3]{}
\providecommand*\mciteSetBstSublistLabelBeginEnd[3]{}
\providecommand*\EndOfBibitem{}
\mciteSetBstSublistMode{f}
\mciteSetBstMaxWidthForm{subitem}{(\alph{mcitesubitemcount})}
\mciteSetBstSublistLabelBeginEnd
  {\mcitemaxwidthsubitemform\space}
  {\relax}
  {\relax}

\bibitem[Baibich \latin{et~al.}(1988)Baibich, Broto, Fert, Van~Dau, Petroff,
  Etienne, Creuzet, Friederich, and Chazelas]{baibich1988giant}
Baibich,~M.~N.; Broto,~J.~M.; Fert,~A.; Van~Dau,~F.~N.; Petroff,~F.;
  Etienne,~P.; Creuzet,~G.; Friederich,~A.; Chazelas,~J. Giant
  magnetoresistance of (001) Fe/(001) Cr magnetic superlattices. \emph{Physical
  review letters} \textbf{1988}, \emph{61}, 2472\relax
\mciteBstWouldAddEndPuncttrue
\mciteSetBstMidEndSepPunct{\mcitedefaultmidpunct}
{\mcitedefaultendpunct}{\mcitedefaultseppunct}\relax
\EndOfBibitem
\bibitem[Binasch \latin{et~al.}(1989)Binasch, Gr{\"u}nberg, Saurenbach, and
  Zinn]{binasch1989enhanced}
Binasch,~G.; Gr{\"u}nberg,~P.; Saurenbach,~F.; Zinn,~W. Enhanced
  magnetoresistance in layered magnetic structures with antiferromagnetic
  interlayer exchange. \emph{Physical review B} \textbf{1989}, \emph{39},
  4828\relax
\mciteBstWouldAddEndPuncttrue
\mciteSetBstMidEndSepPunct{\mcitedefaultmidpunct}
{\mcitedefaultendpunct}{\mcitedefaultseppunct}\relax
\EndOfBibitem
\bibitem[Slonczewski \latin{et~al.}(1996)Slonczewski, \latin{et~al.}
  others]{slonczewski1996current}
Slonczewski,~J.~C., \latin{et~al.}  Current-driven excitation of magnetic
  multilayers. \emph{Journal of Magnetism and Magnetic Materials}
  \textbf{1996}, \emph{159}, L1\relax
\mciteBstWouldAddEndPuncttrue
\mciteSetBstMidEndSepPunct{\mcitedefaultmidpunct}
{\mcitedefaultendpunct}{\mcitedefaultseppunct}\relax
\EndOfBibitem
\bibitem[Myers \latin{et~al.}(1999)Myers, Ralph, Katine, Louie, and
  Buhrman]{myers1999current}
Myers,~E.; Ralph,~D.; Katine,~J.; Louie,~R.; Buhrman,~R. Current-induced
  switching of domains in magnetic multilayer devices. \emph{Science}
  \textbf{1999}, \emph{285}, 867--870\relax
\mciteBstWouldAddEndPuncttrue
\mciteSetBstMidEndSepPunct{\mcitedefaultmidpunct}
{\mcitedefaultendpunct}{\mcitedefaultseppunct}\relax
\EndOfBibitem
\bibitem[Awschalom and Flatt{\'e}(2007)Awschalom, and
  Flatt{\'e}]{awschalom2007challenges}
Awschalom,~D.~D.; Flatt{\'e},~M.~E. Challenges for semiconductor spintronics.
  \emph{Nature physics} \textbf{2007}, \emph{3}, 153--159\relax
\mciteBstWouldAddEndPuncttrue
\mciteSetBstMidEndSepPunct{\mcitedefaultmidpunct}
{\mcitedefaultendpunct}{\mcitedefaultseppunct}\relax
\EndOfBibitem
\bibitem[Gong \latin{et~al.}(2017)Gong, Li, Li, Ji, Stern, Xia, Cao, Bao, Wang,
  Wang, \latin{et~al.} others]{gong2017discovery}
Gong,~C.; Li,~L.; Li,~Z.; Ji,~H.; Stern,~A.; Xia,~Y.; Cao,~T.; Bao,~W.;
  Wang,~C.; Wang,~Y., \latin{et~al.}  Discovery of intrinsic ferromagnetism in
  two-dimensional van der Waals crystals. \emph{Nature} \textbf{2017},
  \emph{546}, 265--269\relax
\mciteBstWouldAddEndPuncttrue
\mciteSetBstMidEndSepPunct{\mcitedefaultmidpunct}
{\mcitedefaultendpunct}{\mcitedefaultseppunct}\relax
\EndOfBibitem
\bibitem[Burch \latin{et~al.}(2018)Burch, Mandrus, and
  Park]{burch2018magnetism}
Burch,~K.~S.; Mandrus,~D.; Park,~J.-G. Magnetism in two-dimensional van der
  Waals materials. \emph{Nature} \textbf{2018}, \emph{563}, 47--52\relax
\mciteBstWouldAddEndPuncttrue
\mciteSetBstMidEndSepPunct{\mcitedefaultmidpunct}
{\mcitedefaultendpunct}{\mcitedefaultseppunct}\relax
\EndOfBibitem
\bibitem[Gong and Zhang(2019)Gong, and Zhang]{gong2019two}
Gong,~C.; Zhang,~X. Two-dimensional magnetic crystals and emergent
  heterostructure devices. \emph{Science} \textbf{2019}, \emph{363},
  eaav4450\relax
\mciteBstWouldAddEndPuncttrue
\mciteSetBstMidEndSepPunct{\mcitedefaultmidpunct}
{\mcitedefaultendpunct}{\mcitedefaultseppunct}\relax
\EndOfBibitem
\bibitem[Geim and Grigorieva(2013)Geim, and Grigorieva]{geim2013van}
Geim,~A.~K.; Grigorieva,~I.~V. Van der Waals heterostructures. \emph{Nature}
  \textbf{2013}, \emph{499}, 419--425\relax
\mciteBstWouldAddEndPuncttrue
\mciteSetBstMidEndSepPunct{\mcitedefaultmidpunct}
{\mcitedefaultendpunct}{\mcitedefaultseppunct}\relax
\EndOfBibitem
\bibitem[Cardoso \latin{et~al.}(2018)Cardoso, Soriano,
  Garc{\'\i}a-Mart{\'\i}nez, and Fern{\'a}ndez-Rossier]{cardoso2018van}
Cardoso,~C.; Soriano,~D.; Garc{\'\i}a-Mart{\'\i}nez,~N.;
  Fern{\'a}ndez-Rossier,~J. Van der Waals spin valves. \emph{Physical review
  letters} \textbf{2018}, \emph{121}, 067701\relax
\mciteBstWouldAddEndPuncttrue
\mciteSetBstMidEndSepPunct{\mcitedefaultmidpunct}
{\mcitedefaultendpunct}{\mcitedefaultseppunct}\relax
\EndOfBibitem
\bibitem[Zhai \latin{et~al.}(2021)Zhai, Xu, Cui, Zhu, Yang, and
  Blanter]{zhai2021electrically}
Zhai,~X.; Xu,~Z.; Cui,~Q.; Zhu,~Y.; Yang,~H.; Blanter,~Y.~M. Electrically
  Controllable Van Der Waals Antiferromagnetic Spin Valve. \emph{Physical
  Review Applied} \textbf{2021}, \emph{16}, 014032\relax
\mciteBstWouldAddEndPuncttrue
\mciteSetBstMidEndSepPunct{\mcitedefaultmidpunct}
{\mcitedefaultendpunct}{\mcitedefaultseppunct}\relax
\EndOfBibitem
\bibitem[Sierra \latin{et~al.}(2021)Sierra, Fabian, Kawakami, Roche, and
  Valenzuela]{sierra2021van}
Sierra,~J.~F.; Fabian,~J.; Kawakami,~R.~K.; Roche,~S.; Valenzuela,~S.~O. Van
  der Waals heterostructures for spintronics and opto-spintronics. \emph{Nature
  Nanotechnology} \textbf{2021}, 1--13\relax
\mciteBstWouldAddEndPuncttrue
\mciteSetBstMidEndSepPunct{\mcitedefaultmidpunct}
{\mcitedefaultendpunct}{\mcitedefaultseppunct}\relax
\EndOfBibitem
\bibitem[Han \latin{et~al.}(2014)Han, Kawakami, Gmitra, and Fabian]{Han2014}
Han,~W.; Kawakami,~R.~K.; Gmitra,~M.; Fabian,~J. Graphene spintronics.
  \emph{Nat Nano} \textbf{2014}, \emph{9}, 794--807, Review\relax
\mciteBstWouldAddEndPuncttrue
\mciteSetBstMidEndSepPunct{\mcitedefaultmidpunct}
{\mcitedefaultendpunct}{\mcitedefaultseppunct}\relax
\EndOfBibitem
\bibitem[Dr\"{o}geler \latin{et~al.}(2016)Dr\"{o}geler, Franzen, Volmer,
  Pohlmann, Banszerus, Wolter, Watanabe, Taniguchi, Stampfer, and
  Beschoten]{drogeler2016spin}
Dr\"{o}geler,~M.; Franzen,~C.; Volmer,~F.; Pohlmann,~T.; Banszerus,~L.;
  Wolter,~M.; Watanabe,~K.; Taniguchi,~T.; Stampfer,~C.; Beschoten,~B. Spin
  lifetimes exceeding 12 ns in graphene nonlocal spin valve devices. \emph{Nano
  letters} \textbf{2016}, \emph{16}, 3533--3539\relax
\mciteBstWouldAddEndPuncttrue
\mciteSetBstMidEndSepPunct{\mcitedefaultmidpunct}
{\mcitedefaultendpunct}{\mcitedefaultseppunct}\relax
\EndOfBibitem
\bibitem[Garcia \latin{et~al.}(2018)Garcia, Vila, Cummings, and
  Roche]{garcia2018spin}
Garcia,~J.~H.; Vila,~M.; Cummings,~A.~W.; Roche,~S. Spin transport in
  graphene/transition metal dichalcogenide heterostructures. \emph{Chemical
  Society Reviews} \textbf{2018}, \emph{47}, 3359--3379\relax
\mciteBstWouldAddEndPuncttrue
\mciteSetBstMidEndSepPunct{\mcitedefaultmidpunct}
{\mcitedefaultendpunct}{\mcitedefaultseppunct}\relax
\EndOfBibitem
\bibitem[Zollner \latin{et~al.}(2020)Zollner, Gmitra, and
  Fabian]{zollner2020swapping}
Zollner,~K.; Gmitra,~M.; Fabian,~J. Swapping exchange and spin-orbit coupling
  in 2d van der waals heterostructures. \emph{Physical Review Letters}
  \textbf{2020}, \emph{125}, 196402\relax
\mciteBstWouldAddEndPuncttrue
\mciteSetBstMidEndSepPunct{\mcitedefaultmidpunct}
{\mcitedefaultendpunct}{\mcitedefaultseppunct}\relax
\EndOfBibitem
\bibitem[Kurebayashi \latin{et~al.}(2022)Kurebayashi, Garcia, Khan, Sinova, and
  Roche]{kurebayashi2022magnetism}
Kurebayashi,~H.; Garcia,~J.~H.; Khan,~S.; Sinova,~J.; Roche,~S. Magnetism,
  symmetry and spin transport in van der Waals layered systems. \emph{Nature
  Reviews Physics} \textbf{2022}, 1--17\relax
\mciteBstWouldAddEndPuncttrue
\mciteSetBstMidEndSepPunct{\mcitedefaultmidpunct}
{\mcitedefaultendpunct}{\mcitedefaultseppunct}\relax
\EndOfBibitem
\bibitem[Safeer \latin{et~al.}(2019)Safeer, Ingla-Ayn{\'e}s, Herling, Garcia,
  Vila, Ontoso, Calvo, Roche, Hueso, and Casanova]{safeer2019room}
Safeer,~C.; Ingla-Ayn{\'e}s,~J.; Herling,~F.; Garcia,~J.~H.; Vila,~M.;
  Ontoso,~N.; Calvo,~M.~R.; Roche,~S.; Hueso,~L.~E.; Casanova,~F.
  Room-Temperature Spin Hall Effect in Graphene/MoS2 van der Waals
  Heterostructures. \emph{Nano letters} \textbf{2019}, \emph{19},
  1074--1082\relax
\mciteBstWouldAddEndPuncttrue
\mciteSetBstMidEndSepPunct{\mcitedefaultmidpunct}
{\mcitedefaultendpunct}{\mcitedefaultseppunct}\relax
\EndOfBibitem
\bibitem[Ghiasi \latin{et~al.}(2019)Ghiasi, Kaverzin, Blah, and van
  Wees]{ghiasi2019charge}
Ghiasi,~T.~S.; Kaverzin,~A.~A.; Blah,~P.~J.; van Wees,~B.~J. Charge-to-Spin
  Conversion by the Rashba--Edelstein Effect in Two-Dimensional van der Waals
  Heterostructures up to Room Temperature. \emph{Nano letters} \textbf{2019},
  \emph{19}, 5959--5966\relax
\mciteBstWouldAddEndPuncttrue
\mciteSetBstMidEndSepPunct{\mcitedefaultmidpunct}
{\mcitedefaultendpunct}{\mcitedefaultseppunct}\relax
\EndOfBibitem
\bibitem[Ben{\'\i}tez \latin{et~al.}(2020)Ben{\'\i}tez, Torres, Sierra,
  Timmermans, Garcia, Roche, Costache, and Valenzuela]{benitez2020tunable}
Ben{\'\i}tez,~L.~A.; Torres,~W.~S.; Sierra,~J.~F.; Timmermans,~M.;
  Garcia,~J.~H.; Roche,~S.; Costache,~M.~V.; Valenzuela,~S.~O. Tunable
  room-temperature spin galvanic and spin Hall effects in van der Waals
  heterostructures. \emph{Nature Materials} \textbf{2020}, 1--6\relax
\mciteBstWouldAddEndPuncttrue
\mciteSetBstMidEndSepPunct{\mcitedefaultmidpunct}
{\mcitedefaultendpunct}{\mcitedefaultseppunct}\relax
\EndOfBibitem
\bibitem[Khokhriakov \latin{et~al.}(2020)Khokhriakov, Hoque, Karpiak, and
  Dash]{khokhriakov2020gate}
Khokhriakov,~D.; Hoque,~A.~M.; Karpiak,~B.; Dash,~S.~P. Gate-tunable
  spin-galvanic effect in graphene-topological insulator van der Waals
  heterostructures at room temperature. \emph{Nature communications}
  \textbf{2020}, \emph{11}, 1--7\relax
\mciteBstWouldAddEndPuncttrue
\mciteSetBstMidEndSepPunct{\mcitedefaultmidpunct}
{\mcitedefaultendpunct}{\mcitedefaultseppunct}\relax
\EndOfBibitem
\bibitem[Wei \latin{et~al.}(2016)Wei, Lee, Lemaitre, Pinel, Cutaia, Cha,
  Katmis, Zhu, Heiman, Hone, \latin{et~al.} others]{wei2016strong}
Wei,~P.; Lee,~S.; Lemaitre,~F.; Pinel,~L.; Cutaia,~D.; Cha,~W.; Katmis,~F.;
  Zhu,~Y.; Heiman,~D.; Hone,~J., \latin{et~al.}  Strong interfacial exchange
  field in the graphene/EuS heterostructure. \emph{Nature materials}
  \textbf{2016}, \emph{15}, 711--716\relax
\mciteBstWouldAddEndPuncttrue
\mciteSetBstMidEndSepPunct{\mcitedefaultmidpunct}
{\mcitedefaultendpunct}{\mcitedefaultseppunct}\relax
\EndOfBibitem
\bibitem[Behera \latin{et~al.}(2019)Behera, Bora, Chowdhury, and
  Deb]{behera2019proximity}
Behera,~S.~K.; Bora,~M.; Chowdhury,~S. S.~P.; Deb,~P. Proximity effects in
  graphene and ferromagnetic CrBr3 van der Waals heterostructures.
  \emph{Physical Chemistry Chemical Physics} \textbf{2019}, \emph{21},
  25788--25796\relax
\mciteBstWouldAddEndPuncttrue
\mciteSetBstMidEndSepPunct{\mcitedefaultmidpunct}
{\mcitedefaultendpunct}{\mcitedefaultseppunct}\relax
\EndOfBibitem
\bibitem[Wang \latin{et~al.}(2015)Wang, Tang, Sachs, Barlas, and
  Shi]{wang2015proximity}
Wang,~Z.; Tang,~C.; Sachs,~R.; Barlas,~Y.; Shi,~J. Proximity-induced
  ferromagnetism in graphene revealed by the anomalous Hall effect.
  \emph{Physical review letters} \textbf{2015}, \emph{114}, 016603\relax
\mciteBstWouldAddEndPuncttrue
\mciteSetBstMidEndSepPunct{\mcitedefaultmidpunct}
{\mcitedefaultendpunct}{\mcitedefaultseppunct}\relax
\EndOfBibitem
\bibitem[Tang \latin{et~al.}(2018)Tang, Cheng, Aldosary, Wang, Jiang, Watanabe,
  Taniguchi, Bockrath, and Shi]{tang2018approaching}
Tang,~C.; Cheng,~B.; Aldosary,~M.; Wang,~Z.; Jiang,~Z.; Watanabe,~K.;
  Taniguchi,~T.; Bockrath,~M.; Shi,~J. Approaching quantum anomalous Hall
  effect in proximity-coupled YIG/graphene/h-BN sandwich structure. \emph{APL
  Materials} \textbf{2018}, \emph{6}, 026401\relax
\mciteBstWouldAddEndPuncttrue
\mciteSetBstMidEndSepPunct{\mcitedefaultmidpunct}
{\mcitedefaultendpunct}{\mcitedefaultseppunct}\relax
\EndOfBibitem
\bibitem[Ghiasi \latin{et~al.}(2021)Ghiasi, Kaverzin, Dismukes, de~Wal, Roy,
  and van Wees]{ghiasi2021electrical}
Ghiasi,~T.~S.; Kaverzin,~A.~A.; Dismukes,~A.~H.; de~Wal,~D.~K.; Roy,~X.; van
  Wees,~B.~J. Electrical and thermal generation of spin currents by magnetic
  bilayer graphene. \emph{Nature nanotechnology} \textbf{2021}, 1--7\relax
\mciteBstWouldAddEndPuncttrue
\mciteSetBstMidEndSepPunct{\mcitedefaultmidpunct}
{\mcitedefaultendpunct}{\mcitedefaultseppunct}\relax
\EndOfBibitem
\bibitem[Ghiasi \latin{et~al.}(2017)Ghiasi, Ingla-Ayn\'{e}s, Kaverzin, and van
  Wees]{ghiasi2017large}
Ghiasi,~T.~S.; Ingla-Ayn\'{e}s,~J.; Kaverzin,~A.~A.; van Wees,~B.~J. Large
  proximity-induced spin lifetime anisotropy in transition-metal
  dichalcogenide/graphene heterostructures. \emph{Nano letters} \textbf{2017},
  \emph{17}, 7528--7532\relax
\mciteBstWouldAddEndPuncttrue
\mciteSetBstMidEndSepPunct{\mcitedefaultmidpunct}
{\mcitedefaultendpunct}{\mcitedefaultseppunct}\relax
\EndOfBibitem
\bibitem[Ben\'{i}tez \latin{et~al.}(2018)Ben\'{i}tez, Sierra, Torres, Arrighi,
  Bonell, Costache, and Valenzuela]{benitez2018strongly}
Ben\'{i}tez,~L.~A.; Sierra,~J.~F.; Torres,~W.~S.; Arrighi,~A.; Bonell,~F.;
  Costache,~M.~V.; Valenzuela,~S.~O. Strongly anisotropic spin relaxation in
  graphene--transition metal dichalcogenide heterostructures at room
  temperature. \emph{Nature Physics} \textbf{2018}, \emph{14}, 303\relax
\mciteBstWouldAddEndPuncttrue
\mciteSetBstMidEndSepPunct{\mcitedefaultmidpunct}
{\mcitedefaultendpunct}{\mcitedefaultseppunct}\relax
\EndOfBibitem
\bibitem[Zihlmann \latin{et~al.}(2019)Zihlmann, Garcia, Watanabe,
  Sch{\"o}nenberger, Makk, Cummings, Kedves, and Taniguchi]{zihlmann2019large}
Zihlmann,~S.; Garcia,~J.~H.; Watanabe,~K.; Sch{\"o}nenberger,~C.; Makk,~P.;
  Cummings,~A.~W.; Kedves,~M.; Taniguchi,~T. Large spin relaxation anisotropy
  and valley-Zeeman spin-orbit coupling in WSe2/Gr/hBN heterostructures.
  \emph{Physical review B} \textbf{2019}, \emph{97}\relax
\mciteBstWouldAddEndPuncttrue
\mciteSetBstMidEndSepPunct{\mcitedefaultmidpunct}
{\mcitedefaultendpunct}{\mcitedefaultseppunct}\relax
\EndOfBibitem
\bibitem[Ingla-Ayn{\'e}s \latin{et~al.}(2021)Ingla-Ayn{\'e}s, Herling, Fabian,
  Hueso, and Casanova]{ingla2021electrical}
Ingla-Ayn{\'e}s,~J.; Herling,~F.; Fabian,~J.; Hueso,~L.~E.; Casanova,~F.
  Electrical Control of Valley-Zeeman Spin-Orbit-Coupling--Induced Spin
  Precession at Room Temperature. \emph{Physical Review Letters} \textbf{2021},
  \emph{127}, 047202\relax
\mciteBstWouldAddEndPuncttrue
\mciteSetBstMidEndSepPunct{\mcitedefaultmidpunct}
{\mcitedefaultendpunct}{\mcitedefaultseppunct}\relax
\EndOfBibitem
\bibitem[Leutenantsmeyer \latin{et~al.}(2016)Leutenantsmeyer, Kaverzin,
  Wojtaszek, and Van~Wees]{leutenantsmeyer2016proximity}
Leutenantsmeyer,~J.~C.; Kaverzin,~A.~A.; Wojtaszek,~M.; Van~Wees,~B.~J.
  Proximity induced room temperature ferromagnetism in graphene probed with
  spin currents. \emph{2D Materials} \textbf{2016}, \emph{4}, 014001\relax
\mciteBstWouldAddEndPuncttrue
\mciteSetBstMidEndSepPunct{\mcitedefaultmidpunct}
{\mcitedefaultendpunct}{\mcitedefaultseppunct}\relax
\EndOfBibitem
\bibitem[Singh \latin{et~al.}(2017)Singh, Katoch, Zhu, Meng, Liu, Brangham,
  Yang, Flatt{\'e}, and Kawakami]{singh2017strong}
Singh,~S.; Katoch,~J.; Zhu,~T.; Meng,~K.-Y.; Liu,~T.; Brangham,~J.~T.;
  Yang,~F.; Flatt{\'e},~M.~E.; Kawakami,~R.~K. Strong modulation of spin
  currents in bilayer graphene by static and fluctuating proximity exchange
  fields. \emph{Physical review letters} \textbf{2017}, \emph{118},
  187201\relax
\mciteBstWouldAddEndPuncttrue
\mciteSetBstMidEndSepPunct{\mcitedefaultmidpunct}
{\mcitedefaultendpunct}{\mcitedefaultseppunct}\relax
\EndOfBibitem
\bibitem[Karpiak \latin{et~al.}(2019)Karpiak, Cummings, Zollner, Vila,
  Khokhriakov, Hoque, Dankert, Svedlindh, Fabian, Roche, \latin{et~al.}
  others]{karpiak2019magnetic}
Karpiak,~B.; Cummings,~A.~W.; Zollner,~K.; Vila,~M.; Khokhriakov,~D.;
  Hoque,~A.~M.; Dankert,~A.; Svedlindh,~P.; Fabian,~J.; Roche,~S.,
  \latin{et~al.}  Magnetic proximity in a van der Waals heterostructure of
  magnetic insulator and graphene. \emph{2D Materials} \textbf{2019}, \emph{7},
  015026\relax
\mciteBstWouldAddEndPuncttrue
\mciteSetBstMidEndSepPunct{\mcitedefaultmidpunct}
{\mcitedefaultendpunct}{\mcitedefaultseppunct}\relax
\EndOfBibitem
\bibitem[Wu \latin{et~al.}(2020)Wu, Yin, Pan, Grutter, Pan, Lee, Gilbert,
  Borchers, Ratcliff, Li, \latin{et~al.} others]{wu2020large}
Wu,~Y.; Yin,~G.; Pan,~L.; Grutter,~A.~J.; Pan,~Q.; Lee,~A.; Gilbert,~D.~A.;
  Borchers,~J.~A.; Ratcliff,~W.; Li,~A., \latin{et~al.}  Large exchange
  splitting in monolayer graphene magnetized by an antiferromagnet.
  \emph{Nature Electronics} \textbf{2020}, \emph{3}, 604--611\relax
\mciteBstWouldAddEndPuncttrue
\mciteSetBstMidEndSepPunct{\mcitedefaultmidpunct}
{\mcitedefaultendpunct}{\mcitedefaultseppunct}\relax
\EndOfBibitem
\bibitem[Telford \latin{et~al.}(2020)Telford, Dismukes, Lee, Cheng, Wieteska,
  Bartholomew, Chen, Xu, Pasupathy, Zhu, \latin{et~al.}
  others]{telford2020layered}
Telford,~E.~J.; Dismukes,~A.~H.; Lee,~K.; Cheng,~M.; Wieteska,~A.;
  Bartholomew,~A.~K.; Chen,~Y.-S.; Xu,~X.; Pasupathy,~A.~N.; Zhu,~X.,
  \latin{et~al.}  Layered antiferromagnetism induces large negative
  magnetoresistance in the van der waals semiconductor CrSBr. \emph{Advanced
  Materials} \textbf{2020}, \emph{32}, 2003240\relax
\mciteBstWouldAddEndPuncttrue
\mciteSetBstMidEndSepPunct{\mcitedefaultmidpunct}
{\mcitedefaultendpunct}{\mcitedefaultseppunct}\relax
\EndOfBibitem
\bibitem[Lee \latin{et~al.}(2021)Lee, Dismukes, Telford, Wiscons, Wang, Xu,
  Nuckolls, Dean, Roy, and Zhu]{lee2021magnetic}
Lee,~K.; Dismukes,~A.~H.; Telford,~E.~J.; Wiscons,~R.~A.; Wang,~J.; Xu,~X.;
  Nuckolls,~C.; Dean,~C.~R.; Roy,~X.; Zhu,~X. Magnetic order and symmetry in
  the 2D semiconductor CrSBr. \emph{Nano Letters} \textbf{2021}, \emph{21},
  3511--3517\relax
\mciteBstWouldAddEndPuncttrue
\mciteSetBstMidEndSepPunct{\mcitedefaultmidpunct}
{\mcitedefaultendpunct}{\mcitedefaultseppunct}\relax
\EndOfBibitem
\bibitem[Zayets(2012)]{zayets2012spin}
Zayets,~V. Spin and charge transport in materials with spin-dependent
  conductivity. \emph{Physical Review B} \textbf{2012}, \emph{86}, 174415\relax
\mciteBstWouldAddEndPuncttrue
\mciteSetBstMidEndSepPunct{\mcitedefaultmidpunct}
{\mcitedefaultendpunct}{\mcitedefaultseppunct}\relax
\EndOfBibitem
\bibitem[Johnson and Silsbee(1985)Johnson, and Silsbee]{johnson1985interfacial}
Johnson,~M.; Silsbee,~R.~H. Interfacial charge-spin coupling: Injection and
  detection of spin magnetization in metals. \emph{Physical review letters}
  \textbf{1985}, \emph{55}, 1790\relax
\mciteBstWouldAddEndPuncttrue
\mciteSetBstMidEndSepPunct{\mcitedefaultmidpunct}
{\mcitedefaultendpunct}{\mcitedefaultseppunct}\relax
\EndOfBibitem
\end{mcitethebibliography}
\end{document}